\documentclass{article}
\usepackage{amsfonts}
\usepackage{makeidx}
\usepackage{amsmath}
\usepackage{amssymb}
\usepackage{graphicx}

\input{tcilatex}

\begin{document}

\title{Three fluid cosmological model using Lie and Noether symmetries}
\author{Michael Tsamparlis\thanks{%
Email: mtsampa@phys.uoa.gr} \ and Andronikos Paliathanasis\thanks{%
Email: anpaliat@phys.uoa.gr} \\
{\small \textit{Faculty of Physics, Department of
Astronomy-Astrophysics-Mechanics,}}\\
{\small \textit{\ University of Athens, Panepistemiopolis, Athens 157 83,
Greece}}}
\date{}
\maketitle

\begin{abstract}
We employ a three fluid model in order to construct a cosmological model in
the Friedmann Robertson Walker flat spacetime, which contains three types of
matter dark energy, dark matter and a perfect fluid with a linear equation
of state. Dark matter is described by dust and dark energy with a scalar
field with potential $V(\phi )$. In order to fix the scalar field potential
we demand Lie symmetry invariance of the field equations, which is a
model-independent assumption. The requirement of an extra Lie symmetry
selects the exponential scalar field potential. The further requirement that
the analytic solution is invariant under the point transformation generated
by the Lie symmetry eliminates dark matter and leads to a quintessence and a
phantom cosmological model containing a perfect fluid and a scalar field.
Next we assume that the Lagrangian of the system admits an extra Noether
symmetry. This new assumption selects the scalar field potential to be
exponential and forces the perfect fluid to be stiff. Furthermore the
existence of the Noether integral allows for the integration of the
dynamical equations. We find new analytic solutions to quintessence and
phantom cosmologies which contain all three fluids. Using these solutions
one is able to compute analytically all main cosmological functions, such as
the scale factor, the scalar field, the Hubble expansion rate, the
deceleration parameter etc.

PACS\ numbers: 98.80.-k, 95.35.+d, 95.36.+x

\end{abstract}

\section{Introduction}

The recent cosmological data indicate that the universe (a) is spatially
flat, (b)\ has suffered two acceleration phases. An early acceleration phase
(inflation), which occurred prior to the radiation dominated era and a
recently initiated accelerated expansion. The source for the late time
cosmic acceleration has been attributed to an unidentified type of matter,
the dark energy (DE). DE contrary to the ordinary baryonic matter has a
negative pressure which counteracts the gravitational force and leads to the
observed accelerated expansion. The nature of DE is still an open question.

The simplest DE probe is the cosmological constant $\Lambda $\ (vacuum)
leading to the $\Lambda$CDM cosmology \cite{Weinberg89,Peebles03,Pad03}.
However, it has been shown that $\Lambda$CDM cosmology suffers from two
major drawbacks known as \ the fine tuning problem and the coincidence
problem \cite{Peri08}. Besides $\Lambda$CDM cosmology, \ many other
candidates have been proposed in the literature, such as time-varying $%
\Lambda (t)$\ cosmologies, quintessence, $k-$essence, tachyons,
modifications of gravity, Chaplygin gas and others \cite%
{Ratra88,Lambda4,Bas09c,Linder2004,LSS08,Brookfield2005td}.

In addition to DE and the ordinary baryonic matter, it is believed that the
Universe contains a third type of matter, the dark matter (DM). This type of
matter is assumed to be pressureless (non-relativistic) and interact very
weakly with the standard baryonic matter. Therefore its presence is mainly
inferred form gravitational effects on visible matter.

In the following we consider a model of the Universe which contains three
types of matter. The DM is modeled by a dust fluid, the DE by a scalar field
and the rest of matter by a perfect fluid. Since a scalar field can be
considered to be a perfect fluid (see below) in effect we model the matter
in the Universe in terms of two perfect fluids and a dust. All three fluids
are assumed to be self-interacting and minimally coupled to gravity.

The problem with the above scenario is that there does not exist an
underline principle which will specify uniquely the potential $V(\phi )$ of
the scalar field. Indeed in the literature one finds several potentials $%
V(\phi )$ such as exponential, power law, hyperbolic etc. Therefore it is
not possible that one could find an analytic solution of this model even if
an EoS has been assumed.

The aim of the present work is twofold (a)\ to propose a geometric principle
(`selection rule')\ for specifying the potential $V(\phi )$ and (b)\ to
solve analytically the system of the resulting field equations. Concerning
(a), following a recent paper \cite{Basilakos}, we propose that the
potential should be selected by the requirement that the dynamical system of
the three fluids admits an additional Lie or Noether symmetry. This point of
view has also been considered in \cite{Cap09,deRitis,RubanoSFQ,YiZhang}.

The main reason for the consideration of this hypothesis is that the
Lie/Noether point symmetries provide first/ Noether integrals, which assist
the integrability of the system. A fundamental approach to derive the Lie
point and the Noether symmetries of a given dynamical system moving in a
Riemannian space has been proposed recently in \cite{Tsam10}. A similar
analysis can be found in \cite{Olver,StephaniB,Aminova
1995,HaasG,LeachKarasu}.

Concerning the analytic solutions we restrict our considerations to a flat
FRW spacetime. Although there is a great number of papers devoted to the
dynamics of scalar field minimally coupled to matter, not much is known
about analytic exact solutions of these models. Most of the solutions
correspond to spatially flat FRW models with no other source but the scalar
field \cite{Russo,Bertacca,MuslimovSF,MendezExactSF,EllisExactSF}. Recently
in \cite{Basilakos} the analytic solution for matter in the form of dust and
a scalar field has been given. A solution with two scalar fields (where one
of them is a kination i.e. stiff matter) and an exponential potential is
given in \cite{ChimentoTwoSF}.

Few exact solutions are known with spatial curvature \cite%
{HalliwellSF,Easther}. Even less solutions are known for a perfect fluid and
a scalar field \cite{Ratra88,Barrow,ChimentoLPC,ChimentoJacubi}. In
particular in \cite{ChimentoJacubi} the authors consider a prefect fluid
minimally interacting with a scalar field having an arbitrary potential $%
V(\phi )$ and, by considering as variable the scale factor $a,$ they develop
an approach which replaces the potential $V(\phi )$ with an auxiliary
arbitrary function ($F(a)$ in their notation) and a constant ($C$) and give
the implicit solution of the problem. They consider an arbitrary
specification of the function $F(a)$ and the constant $C,$ which leads to an
exponential potential and by (making some further assumptions) they find a
set of analytic solutions in terms of the adiabatic index $\gamma $ of the
fluid. In a different approach to the same problem in \cite{Cap09}, the
authors employ ad hoc an exponential potential and assume that the resulting
system of equations admits a Noether symmetry. They are not able to find an
analytic solution and continue with numerical solutions. This is not
necessary according to the results of \cite{ChimentoJacubi} and our results
below.

In \cite{CapNojiri,NojiriOdi} the authors in an attempt to unify the early
time (phantom) inflation with the late time (phantom or not) acceleration
derive specific solutions within which both the early and the late phases of
evolution of the universe occur. The proposed scenario is based in the
introduction of an additional unspecified function $\omega (\phi )$\ in the
kinetic term of the scalar field Lagrangian. Although this function can be
absorbed into the field during inflation and acceleration by a redefinition
of the scalar field, this is not possible right on the transition point, at
which this term connects the two phases of evolution smoothly. A\ similar
scenario considers two scalar fields.

There does not appear to exist a solution concerning a perfect fluid, a dust
and a scalar field all minimally coupled to gravity and non-interacting in a
spatially flat FRW spacetime. In the following using symmetry assumptions
only, we obtain many of the above solutions and also a new solution which
involves all three fluids. This solution contains the solutions of \cite%
{ChimentoTwoSF} as a special case.

The structure of the paper is as follows. The basic equations and the
resulting dynamical system are presented in section 2. In section 3 we
review briefly the basics of the theory of Lie and Noether symmetries. In
section 4 we define the scalar field potential and subsequently we determine
analytic solutions of the field equations using an extra Lie symmetry and
the Lie invariance of the solutions with respect to this symmetry. In
section 5 we determine a new solution of the three fluid dynamical system
assuming only an extra Noether symmetry. The main conclusions are summarized
in section 6.

\section{The dynamical system}

We consider the flat FRW spacetime in canonical coordinates:\qquad\ 
\begin{equation}
ds^{2}=-dt^{2}+a^{2}\left( t\right) \left( dx^{2}+dy^{2}+dz^{2}\right)
\label{StM.01}
\end{equation}%
with comoving observers $u^{a}=\delta _{t}^{a}$. We also consider that
spacetime contains dark energy, dark matter and `other matter'. DM is
assumed to be described by dust ($p_{DM}=0)$ with energy momentum tensor 
\begin{equation}
_{DM}T_{ab}=\mu _{DM}u_{a}u_{b}.
\end{equation}%
The `other matter' is assumed to be described by a perfect fluid with energy
momentum tensor%
\begin{equation}
_{PF}T_{ab}=\mu _{B}u_{a}u_{b}+p_{B}h_{ab}~
\end{equation}%
and a linear EoS%
\begin{equation}
p_{B}=\left( \gamma -1\right) \mu _{B}~,~1<\gamma \leq 2  \label{StM.01.1}
\end{equation}%
where $\gamma $ is the adiabatic index of the fluid and $%
h_{ab}=g_{ab}+u_{a}u_{b}$ is the tensor projecting normal to the
four-velocity $u^{a}$. For $\gamma =1$ the fluid is dust; for $\gamma =2$ is
stiff matter and for $\gamma =\frac{4}{3}$ is a radiation fluid.

Finally DE\ is described by a scalar field $\phi $ (quintessence or phantom)
rolling down a potential $V(\phi )$ whose energy momentum tensor is 
\begin{equation}
_{\phi }T_{ab}=\left( \frac{1}{2}\varepsilon \dot{\phi}^{2}+V(\phi )\right)
u_{a}u_{b}+\left( \frac{1}{2}\varepsilon \dot{\phi}^{2}-V(\phi )\right)
h_{ab}.
\end{equation}

When $\varepsilon =+1$ we have quintessence (real $\phi $) and when $%
\varepsilon =-1$ we have a phantom field (imaginary $\phi $). It follows
that (for the comoving observers)\ the scalar field may be considered as a
perfect fluid with energy density and isotropic pressure:%
\begin{align}
\mu _{\phi }& =\frac{1}{2}\varepsilon \dot{\phi}^{2}+V(\phi )
\label{StM.04a} \\
p_{\phi }& =\frac{1}{2}\varepsilon \dot{\phi}^{2}-V(\phi ).  \label{StM.04b}
\end{align}

Equations (\ref{StM.04a}),(\ref{StM.04b}) imply the EoS:%
\begin{equation}
\mu _{\phi }-p_{\phi }=2V(\phi ).
\end{equation}%
from which follows that the definition/specification of a potential function
for the scalar field $\phi (t)$ essentially defines an EoS for the perfect
fluid defined by the scalar field. It also follows that stiff matter may be
considered as a homogeneous, massless, free (i.e. with vanishing potential)
scalar field minimally coupled to gravity. Because $\frac{p_{B}}{\mu _{B}}=1$
this scalar field is a kination.

All three fluids are assumed to interact minimally, which implies the three
independent conservation equations:%
\begin{equation}
_{DM}T_{\text{ };b}^{ab}=0,\quad _{PF}T_{\text{ };b}^{ab}=0,\quad _{\phi }T_{%
\text{ };b}^{ab}=0.  \label{StM.04e}
\end{equation}

There are six unknowns in the problem i.e. the $a(t),\mu _{B},$ $p_{B},\mu
_{DM},\phi (t),$ $V(\phi )\ $and five field equations i.e. Einstein's
equation:%
\begin{equation}
G_{ab}=_{\phi }T_{ab}+_{DM}T_{ab}+_{PF}T_{ab}
\end{equation}%
the three conservation equations (\ref{StM.04e}) (equation $_{\phi }T_{\text{
};b}^{ab}=0$ is the Klein Gordon equation) and the EoS (\ref{StM.01.1}). We
need one more equation which must be given by an independent assumption.

However there is not a general consensus or accepted practice as to the
nature of this assumption. In the literature one finds the following types
of assumptions/equations:\newline
(i) Equations which are based on the dynamical system approach. That is, in
the case the field equations follow from a Lagrangian one demands that the
Lagrangian admits a Noether symmetry. This assumption results in an extra
Noether integral, which provides the required extra equation. With this
approach there have been given the analytic solutions for various probes
(UCDM, exponential) \cite{Basilakos,Cap09,RubanoSFQ}, and in other models
e.g. modified gravity \cite%
{Cap97,Sanyal,Cap07,RoshanPalatiniNoether,BonannoGLNoether,VakiliFr,Motavali,MahomedTF}%
. An alternative but similar approach is to demand that the field equations
admit an extra Lie (not necessarily a Noether) symmetry and require that the
solution is invariant under the action of the Lie symmetry. (ii) Equations
which are defined by an ansatz on the scale factor \cite%
{MuslimovSF,MendezExactSF,EllisExactSF}. (iii) Equations which are ad hoc
statements on the scalar field potential.

In the present paper we follow the dynamical system approach and demand that
the system admits an extra Lie symmetry and subsequently an extra Noether
symmetry. The Lie symmetry requirement specifies the potential to be
exponential and produces (with a additional invariance assumption) the
perfect fluid solutions found in \cite{ChimentoJacubi}\ and the empty space
solution of \cite{MuslimovSF,EllisExactSF}. The requirement of Noether
symmetry in addition of fixing the potential, it constraints the perfect
fluid to be stiff and determines the analytic solution of the three fluid
problem.

\subsection{The field equations}

Einstein field equations for the metric (\ref{StM.01}), comoving observers
and a perfect fluid of energy density $\mu _{B}$ and isotropic pressure $%
p_{B}$ minimally coupled to a scalar field $\phi $ with scalar field
potential $V(\phi )$ and DM (dust) with energy density $\mu _{DM}$ are:%
\begin{align}
H^{2}& =\frac{k}{3}\left( \mu _{DM}+\mu _{B}+\mu _{\phi }\right) ~\text{%
(Hubble equation)}  \label{StM.02} \\
\frac{\ddot{a}}{a}+\frac{1}{2}H^{2}& =-\frac{k}{2}\left( p_{B}+p_{\phi
}\right) \text{ (deceleration equation)}  \label{StM.03}
\end{align}%
where $k=8\pi G$ is Einstein's gravitational constant. The deceleration
equation (\ref{StM.03}) can be written in the alternative forms%
\begin{equation}
\dot{H}+\frac{3}{2}H^{2}=-\frac{k}{2}\left( p_{B}+p_{\phi }\right)
\label{StM.12a}
\end{equation}%
\begin{equation}
\dot{H}+\frac{k}{2}\left( \mu _{B}+p_{B}+\varepsilon \dot{\phi}^{2}\right)
=0.  \label{StM.12b}
\end{equation}%
Equations (\ref{StM.02}), (\ref{StM.03}) are supplemented by the Klein
Gordon equation 
\begin{equation}
\ddot{\phi}=-3H\dot{\phi}-\varepsilon V_{,\phi }  \label{StM.12c}
\end{equation}%
and the conservation equations:%
\begin{equation}
\dot{\mu}_{B}+3\left( \mu _{B}+p_{B}\right) H=0  \label{StM.04c}
\end{equation}%
\begin{equation}
\dot{\mu}_{DB}+3\mu _{DB}H=0.  \label{StM.04m}
\end{equation}

The conservation equation (\ref{StM.04c}) and the EoS (\ref{StM.01.1}) imply 
\begin{equation}
\mu _{B}=Ca^{-3\gamma }~,~p_{B}=\left( \gamma -1\right) Ca^{-3\gamma }~,%
\text{ }1<\gamma \leq 2  \label{StM.06}
\end{equation}%
where $C\geqq 0\ $is an integration constant. For DM the conservation
equation (\ref{StM.04m}) gives:%
\begin{equation}
\mu _{DM}=\frac{E}{a^{3}}  \label{StM.06a}
\end{equation}%
where $E\ $is an integration constant. Replacing these results in the field
equations (\ref{StM.02}), (\ref{StM.03}) , (\ref{StM.12c}) we obtain the
system of equations 
\begin{equation}
3a\dot{a}^{2}-\frac{1}{2}\varepsilon ka^{2}\dot{\phi}^{2}-ka^{3}V\left( \phi
\right) -kCa^{-3\left( \gamma -1\right) }=\bar{E}~  \label{StM.10}
\end{equation}%
\begin{align}
\ddot{a}+\frac{1}{2a}\dot{a}^{2}+\frac{k}{4}\varepsilon a\dot{\phi}^{2}-%
\frac{k}{2}aV+\frac{k}{2}\left( \gamma -1\right) Ca^{1-3\gamma }& =0
\label{StM.11} \\
\ddot{\phi}+\frac{3}{a}\dot{a}\dot{\phi}+\varepsilon V_{,\phi }& =0
\label{StM.12}
\end{align}%
where $\bar{E}=kE.$ Equation (\ref{StM.10}) is written in the equivalent form%
\begin{equation}
H^{2}=\frac{k}{3}\left( Ea^{-3}+Ca^{-3\gamma }+\frac{1}{2}\varepsilon \dot{%
\phi}^{2}+V(\phi )\right) .  \label{SFD.7}
\end{equation}

\subsection{Looking for analytic solutions}

In order to find an analytic solution of the system of equations (\ref%
{StM.10}) - (\ref{StM.12}) we need either two independent assumptions, which
will result in two independent conditions, or one stronger assumption which
will result in two conditions. In the following we formulate these
assumptions in terms of Lie and Noether symmetries.

Assuming the unknown potential $V(\phi )$ to be an externally specified
function, the differential equations (\ref{StM.11}), (\ref{StM.12}) contain
only the unknowns $a(t),$ $\phi (t)$. Hence it is possible to consider a new
autonomous two dimensional dynamical system with variables $\{a(t),\phi
(t)\} $ defined by equations (\ref{StM.11}), (\ref{StM.12}) and treat the
third field equation (\ref{StM.10}) (equivalently (\ref{SFD.7})) as a
constraint. Following this remark we look for a Lagrangian producing
equations (\ref{StM.11}), (\ref{StM.12}). It is easily seen that this
Lagrangian is%
\begin{equation}
L=\left( 3a\dot{a}^{2}-\frac{1}{2}k\varepsilon a^{3}\dot{\phi}^{2}\right)
+ka^{3}V\left( \phi \right) +kCa^{-3\left( \gamma -1\right) }  \label{StM.13}
\end{equation}%
with Hamiltonian 
\begin{equation}
\bar{E}=3a\dot{a}^{2}-\frac{1}{2}k\varepsilon a^{3}\dot{\phi}%
^{2}-ka^{3}V\left( \phi \right) -kCa^{-3\left( \gamma -1\right) }.
\label{StM.14}
\end{equation}

We note that the Hamiltonian is the constraint condition (\ref{StM.10}),
that is $\bar{E}=const.$ It is possible to give a geometric interpretation
of the Hamiltonian $\bar{E}.$ Indeed the dynamical system defined by the
Lagrangian (\ref{StM.13}) is autonomous hence admits the Noether symmetry $%
\partial _{t}$ whose Noether integral is precisely the Hamiltonian. We note
that $\bar{E}=0$ only when DM is absent. We arrive at the conclusion:

\emph{The dynamical system defined by the three fluids considered above is
equivalent to a new autonomous two dimensional dynamical system } $%
\{a(t),\phi (t)\}$ \emph{with Lagrangian (\ref{StM.13}) whose Hamiltonian }$%
\bar{E}$ \emph{is a constant which vanishes iff }$\mu _{DM}=0$\emph{.}

In the following we use the Lie and the Noether symmetries of the two
dimensional dynamical system in order to produce analytic solutions.

\section{Lie and Noether symmetries}

Before we proceed we review briefly the basic definitions concerning Lie and
Noether symmetries of systems of second order ordinary differential
equations (ODEs)%
\begin{equation}
\ddot{x}^{i}=\omega ^{i}\left( t,x^{j},\dot{x}^{j}\right) .  \label{Lie.0}
\end{equation}

A\ vector field $X=\xi \left( t,x^{j}\right) \partial _{t}+\eta ^{i}\left(
t,x\right) \partial _{i}$\ in the augmented space $\{t,x^{i}\}$ is the
generator of a Lie point symmetry of the system of ODEs (\ref{Lie.0})\ if
the following condition is satisfied \cite{Olver,StephaniB} 
\begin{equation}
X^{\left[ 2\right] }\left( \ddot{x}^{i}-\omega \left( t,x^{j},\dot{x}%
^{j}\right) \right) =0  \label{Lie.1}
\end{equation}%
where $X^{\left[ 2\right] }$\ is the second prolongation of $X$ defined by
the formula%
\begin{equation}
X^{\left[ 2\right] }=\xi \partial _{t}+\eta ^{i}\partial _{i}+\left( \dot{%
\eta}^{i}-\dot{x}^{i}\dot{\xi}\right) \partial _{\dot{x}^{i}}+\left( \ddot{%
\eta}^{i}-\dot{x}^{i}\ddot{\xi}-2\ddot{x}^{i}\dot{\xi}\right) \partial _{%
\ddot{x}^{i}}.  \label{Lie.2}
\end{equation}%
Condition (\ref{Lie.1}) is equivalent to the relation: 
\begin{equation}
\left[ X^{\left[ 1\right] },A\right] =\lambda \left( x^{a}\right) A
\label{Lie.3a}
\end{equation}%
where $X^{\left[ 1\right] }~$is the first prolongation of $X$ and $A$ is the
Hamiltonian vector field:%
\begin{equation}
A=\partial _{t}+\dot{x}\partial _{x}+\omega ^{i}\left( t,x^{j},\dot{x}%
^{j}\right) \partial _{\dot{x}^{i}}.  \label{Lie.4}
\end{equation}

If the system of ODEs results from a first order Lagrangian $L=L\left(
t,x^{j},\dot{x}^{j}\right) ,$ then a Lie symmetry $X$ of the system is a
Noether symmetry of the Lagrangian if the additional condition is satisfied 
\begin{equation}
X^{\left[ 1\right] }L+L\frac{d\xi }{dt}=\frac{df}{dt}  \label{Lie.5}
\end{equation}%
where $f=f\left( t,x^{j}\right) $\ is the gauge function. To every Noether
symmetry there corresponds a first integral (a Noether integral) of the
system of equations (\ref{Lie.0}) which is given by the formula:%
\begin{equation}
I=\xi E_{H}-\frac{\partial L}{\partial \dot{x}^{i}}\eta ^{i}+f  \label{Lie.6}
\end{equation}%
where $E_{H}$ is the Hamiltonian of $L$.

The vector field $X$ for the Lagrangian (\ref{StM.13}) is 
\begin{equation}
X=\xi \left( t,a,\phi \right) \partial _{t}+\eta _{a}\left( t,a,\phi \right)
\partial _{a}+\eta _{\phi }\left( t,a,\phi \right) \partial _{\phi }
\label{Lie.3}
\end{equation}%
and the first prolongation%
\begin{equation}
X^{\left[ 1\right] }=\xi \partial _{t}+\eta _{a}\partial _{a}+\eta _{\phi
}\partial _{\phi }+\left( \dot{\eta}_{a}-\dot{a}\dot{\xi}\right) \partial _{%
\dot{a}}+\left( \dot{\eta}_{\phi }-\dot{\phi}\dot{\xi}\right) \partial _{%
\dot{\phi}}.  \label{Lie.7}
\end{equation}

Having given the basic formulae for the Lie and Noether symmetries we look
for analytic solutions of the system of equations (\ref{StM.10}) - (\ref%
{StM.12}) using Lie and Noether symmetries.

To simplify the calculations we introduce a new variable $r$ with the
relation $a=r^{\frac{2}{3}}$. In the new variables $r,\phi $ the Lagrangian (%
\ref{StM.13}) takes the form 
\begin{equation}
L=\frac{1}{2}\left( \frac{8}{3}\dot{r}^{2}-\varepsilon kr^{2}\dot{\phi}%
^{2}\right) +kr^{2}V\left( \phi \right) +kCr^{-2\left( \gamma -1\right) }.
\label{StM.16}
\end{equation}

For a general $V(\phi )$ this Lagrangian admits only the standard
Lie/Noether symmetry $\partial _{t}.$ We are looking for the possibility of
extra Lie and Noether symmetries for special forms of the potential $V(\phi
).$ To find these potentials we apply the results of \cite{Tsam10}, which
concern all autonomous two dimensional dynamical systems moving in flat
(Euclidian or Minkowskian) space and admit Lie and/or Noether symmetries.

In order to apply the results of \cite{Tsam10} we need a flat metric and a
potential function for the dynamical system. This is done by decomposing the
Lagrangian (\ref{StM.16}) into kinetic energy, which defines the kinetic
energy metric, and potential energy which defines the potential. We consider
the metric to be:%
\begin{equation}
ds_{L}^{2}=\frac{8}{3}\dot{r}^{2}-\varepsilon kr^{2}\dot{\phi}^{2}
\label{StM.16.1}
\end{equation}%
and the potential function:%
\begin{equation}
W(r,\phi )=-kr^{2}V\left( \phi \right) -kCr^{-2\left( \gamma -1\right) }.
\label{StM.16.2}
\end{equation}

It can be shown that the Ricci scalar of the metric $ds_{L}^{2}$ vanishes,
hence the space $\{r$,$\phi \}$ is the $M^{2}$ (Minkowski two dimensional
space in polar coordinates). Therefore the results of \cite{Tsam10} apply
and we make use of the tables in that paper to read the appropriate
potentials.

\section{Lie symmetry}

We start with the demand that the system of equations (\ref{StM.11}), (\ref%
{StM.12}) admits an extra Lie symmetry (apart of $\partial _{t}$). From
Table 10 Line 5 of \cite{Tsam10} we find that the system of equations admits
an extra Lie symmetry for the potential $V\left( \phi \right)
=V_{0}e^{-d\phi }$ where $~V_{0}~,d$ are constants with $d\neq 0;~$%
Furthermore the Lie symmetry vector is 
\begin{equation}
X=\gamma t\partial _{t}+r\partial _{r}+\frac{2\gamma }{d}\partial _{\phi }
\label{StM.16a}
\end{equation}%
or, in the original coordinates $\alpha (t),\phi \left( t\right) $:%
\begin{equation}
X=\gamma t\partial _{t}+\frac{2}{3}a\partial _{a}+\frac{2\gamma }{d}\partial
_{\phi }.  \label{StM.16.4}
\end{equation}

For the scalar field potential $V(\phi )=V_{0}e^{-d\phi }$ the field
equations become: 
\begin{align}
\left( 3a\dot{a}^{2}-\frac{1}{2}k\varepsilon a^{3}\dot{\phi}^{2}\right)
-ka^{3}V\left( \phi \right) -kCa^{-3\left( \gamma -1\right) }& =\bar{E}
\label{SFD.2} \\
\ddot{a}+\frac{1}{2a}\dot{a}^{2}+\frac{k}{4}\varepsilon a\dot{\phi}^{2}-%
\frac{k}{2}aV_{0}e^{-d\phi }+\frac{k}{2}\left( \gamma -1\right)
Ca^{1-3\gamma }& =0  \label{SFD.3} \\
\ddot{\phi}+3\frac{\dot{a}}{a}\dot{\phi}-d\varepsilon V_{0}e^{-d\phi }& =0.
\label{SFD.4}
\end{align}%
where%
\begin{equation}
\mu _{B}=Ca^{-3\gamma }~,~p_{B}=\left( \gamma -1\right) Ca^{-3\gamma }.
\label{SFD.1}
\end{equation}%
\begin{equation}
\mu _{\phi }=\frac{1}{2}\varepsilon \dot{\phi}^{2}+V_{0}e^{-d\phi
}~,~p_{\phi }=\frac{1}{2}\varepsilon \dot{\phi}^{2}-V_{0}e^{-d\phi }.
\label{SFD.6}
\end{equation}

We look for analytic solutions of these equations.

\subsection{Solution using Lie invariance}

To find a solution of the system of equations we need one further condition.
We require the solution to be invariant under the action of the extra Lie
symmetry (\ref{StM.16.4}). To find this solution we compute the zeroth order
invariants of $X$ using the associated Lagrange system (recall that $%
1<\gamma \leq 2~$and $d\neq 0$):%
\begin{equation*}
\frac{dt}{\gamma t}=\frac{da}{\frac{2}{3}a}=\frac{d\phi }{\frac{2\gamma }{d}}%
.
\end{equation*}%
The solution of the system is the two parameter family of functions: 
\begin{equation}
a(t)=t^{\frac{2}{3\gamma }}~,~\phi (t)=\frac{2}{d}\ln t.  \label{StM.17}
\end{equation}

Subsequently we demand that the (invariant) $a(t),$ $\phi (t)$ we computed
are solutions of the field equations (\ref{StM.11}), (\ref{StM.12}). This
fixes the constants $V_{0},C$ in terms of $\gamma ,d$ as follows:

\begin{equation}
V_{0}=\frac{2\varepsilon \left( 2-\gamma \right) }{\gamma d^{2}}~,C=\frac{%
4\left( d^{2}-3\varepsilon k\gamma \right) }{3k\gamma ^{2}d^{2}}~~,~~\bar{E}%
=0.~  \label{StM.17a.1}
\end{equation}

We note that in both solutions the DM is eliminated because $\bar{E}=0$.
Concerning the properties of these solutions we have the following.

A. In the quintessence case the scalar field $\phi $ is real. Because $%
\gamma \leq 2,$ $V_{0}>0$ the $V(\phi )>0,$ therefore the potential is
"repulsive". The constant $C$ is related to the matter energy $\mu _{B}$ via
relation (\ref{SFD.1}). For the perfect fluid $\mu _{B}$ and $C$ are
positive which requires that $d^{2}>3k\gamma $.

B. In the case of phantom field\ $\varepsilon =-1,$ $\phi $ is complex, $d$
is complex but $V\left( \phi \right) $ is real. Furthermore $V_{0}>0$, and \ 
$C\geqq 0$ provided $\left\vert d^{2}\right\vert \geqq 3k\gamma $. \ 

When $\left\vert d^{2}\right\vert =3k\gamma $ then $C=0$ and we have the
empty space solution with either a quintessence ($\varepsilon =+1$) or a
phantom field ($\varepsilon =-1$). This solution has only the unspecified
parameter $\gamma .$ We note that when $\gamma =2$ (stiff matter) the
constant $V_{0}=0$ which is not acceptable. Therefore the case $\gamma =2$
must be treated separately.

Perfect fluid solutions with a linear EoS and a scalar field with
exponential potential have been considered previously in \cite%
{ChimentoJacubi}.{\LARGE \ }It can be shown that our solution contains the
solution of \cite{ChimentoJacubi}. It is to be noted that we have derived
the solution using only fundamental symmetry assumptions and not ad hoc
statements. Furthermore we have covered the case of the phantom field.

For all cases with $\gamma <2$ we have a genuine perfect fluid solution
minimally interacting with a scalar field . At late time both $\mu
_{B},p_{B}\rightarrow 0,$ that is, the solution (\ref{StM.17}) for late time
tends to the empty space solution with a scalar field. This results agree
with known results (see \cite{Russo,MuslimovSF,EllisExactSF}).

\section{Noether symmetry}

In an alternative approach we continue a stronger symmetry assumption and
require that the dynamical system defined by the Lagrangian (\ref{StM.16})
admits a Noether symmetry. It is known that, contrary to the Lie symmetry,
the Noether symmetry gives two conditions on the system of ODEs which
suffice to produce the analytic solution of the dynamical system

Because the Noether symmetries are Lie symmetries and we have only one extra
Lie symmetry, the Noether symmetry must coincide with the Lie symmetry (\ref%
{StM.16.4}) we found above. This leads to the value of the parameter~$\gamma
=2$. Therefore the requirement of Noether symmetry has the following
implications: (a) Selects the EoS $p_{B}=\mu _{B},$ (b) the original field
equations (\ref{StM.11}), (\ref{StM.12}) are non-homogeneous. (c) Selects
the scalar field potential $V(\phi )=V_{0}e^{-d\phi }$ where $V_{0}$ is a
constant, (d) fixes the Noether symmetry vector (\ref{StM.16a}) and (e)
Provides the Noether integral:%
\begin{equation}
I_{2}=2t\bar{E}-4a^{2}\dot{a}+\frac{4}{d}\varepsilon ka^{3}\dot{\phi}.
\label{StM.26}
\end{equation}%
In the original coordinates $a,\phi $ we have

Lagrangian:%
\begin{equation}
L=\left( 3a\dot{a}^{2}-\frac{1}{2}k\varepsilon a^{3}\dot{\phi}^{2}\right)
+kV_{0}a^{3}e^{-d\phi }+k\bar{C}a^{-3}  \label{StM.24a}
\end{equation}

Hamiltonian: 
\begin{equation}
\bar{E}=\left( 3a\dot{a}^{2}-\frac{1}{2}k\varepsilon a^{3}\dot{\phi}%
^{2}\right) -ka^{3}V_{0}e^{-d\phi }-k\bar{C}a^{-3}  \label{StM.25}
\end{equation}

Equations of motion:%
\begin{align}
\ddot{a}+\frac{1}{2a}\dot{a}^{2}+\frac{k}{4}\varepsilon a\dot{\phi}^{2}-%
\frac{k}{2}aV_{0}e^{-d\phi }+\frac{k}{2}\bar{C}a^{-5}& =0  \label{StM.22} \\
\ddot{\phi}+\frac{3}{a}\dot{a}\dot{\phi}-d\varepsilon V_{0}e^{-d\phi }& =0.
\label{StM.23}
\end{align}

We conclude that the requirement of the existence of an extra Noether
symmetry results in a model consisting of a kination, a scalar field with an
exponential potential and dust.

\subsection{The analytic solution}

We determine the analytic solution of the system of field equations (\ref%
{StM.22}), (\ref{StM.23}) under the constraint (\ref{StM.25}). In the
solution we follow a procedure initiated in \cite{Russo}. We introduce new
variables $\{u,v\}$ with the relations%
\begin{align}
u& =\frac{\sqrt{6k\varepsilon }}{4}\phi +\frac{1}{2}\ln \left( a^{3}\right) ~
\label{StM.27} \\
v& =-\frac{\sqrt{6k\varepsilon }}{4}\phi +\frac{1}{2}\ln \left( a^{3}\right)
.~  \label{StM.28}
\end{align}
In the new variables the Lagrangian (\ref{StM.24a}) is 
\begin{equation}
L=e^{\left( u+v\right) }\left[ \frac{4}{3}\dot{u}\dot{v}+\bar{V}%
_{0}e^{-2K\left( u-v\right) }+Ce^{-2\left( u+v\right) }\right]
\label{StM.29}
\end{equation}%
where $K=\frac{d}{\sqrt{6k\varepsilon }},~\bar{V}_{0}=kV_{0}.~$ Next we
consider a change in the time coordinate 
\begin{equation}
\frac{d\tau }{dt}=\sqrt{\frac{3kV_{0}}{4}}e^{-K\left( u-v\right) }.
\label{StM.30}
\end{equation}

Using the variation integral$~$the Lagrangian becomes 
\begin{equation}
L=e^{(1-K)u}e^{\left( 1+K\right) v}\left( u^{\prime }v^{\prime }+1+\bar{C}%
e^{-2(1-K)u}e^{-2\left( 1+K\right) v}\right)
\end{equation}%
where $\bar{C}=\frac{4}{3kV_{0}}C~$and the Hamiltonian 
\begin{equation}
\bar{E}=e^{(1-K)u}e^{\left( 1+K\right) v}\left( u^{\prime }v^{\prime }-1-%
\bar{C}e^{-2(1-K)u}e^{-2\left( 1+K\right) v}\right) .  \label{StM.36b}
\end{equation}
The resulting equations of motion are:%
\begin{align}
u^{\prime \prime }+\left( 1-K\right) u^{\prime 2}-\left( 1+K\right) \left( 1-%
\bar{C}e^{-2\left( 1-K\right) u}e^{-2\left( 1+K\right) v}\right) & =0
\label{StM.35} \\
v^{\prime \prime }+\left( 1+K\right) v^{\prime 2}-\left( 1-K\right) \left( 1-%
\bar{C}e^{-2\left( 1-K\right) u}e^{-2\left( 1+K\right) v}\right) & =0.
\label{StM.36}
\end{align}
We consider the cases $K=1$ and $K\neq 1.$

\subsubsection{The case $K=1$}

For $K=1$ the system of equations (\ref{StM.35}), (\ref{StM.36}) becomes:

\begin{align}
u^{\prime \prime }-2\left( 1-\bar{C}e^{-4v}\right) & =0  \label{StM.37} \\
v^{\prime \prime }+2v^{\prime 2}& =0  \label{StM.38}
\end{align}%
and the Hamiltonian (\ref{StM.36b})%
\begin{equation}
\bar{E}=e^{2v}\left( u^{\prime }v^{\prime }-1-\bar{C}e^{-4v}\right) .
\label{StM.39}
\end{equation}
The solution of the system (\ref{StM.37}) - (\ref{StM.39}) is 
\begin{align}
u\left( \tau \right) & =\tau ^{2}+\frac{1}{2}\bar{C}\ln \left( \tau
+c\right) +\left( \bar{E}+2c\right) \tau  \label{StM.40} \\
v\left( \tau \right) & =\frac{1}{2}\ln \left( 2\tau +2c\right) .
\label{StM.41}
\end{align}
Replacing in (\ref{StM.27}),(\ref{StM.28}) and (\ref{StM.29}) we find the
analytic solution%
\begin{equation}
ds^{2}=-N^{2}\left( \tau \right) d\tau ^{2}+a^{2}\left( \tau \right) \delta
_{ij}dx^{i}dx^{j}  \label{StM.43}
\end{equation}%
where%
\begin{align}
a^{3}\left( \tau \right) & =\sqrt{2\left( \tau +c\right) ^{\bar{C}}\left(
\tau +c\right) }e^{\tau ^{2}+\left( \bar{E}+2c\right) \tau }  \label{StM.45}
\\
\phi \left( \tau \right) & =\frac{2}{\sqrt{6k\varepsilon }}\left[ \tau
^{2}+\left( \bar{E}+2c\right) \tau +\frac{1}{2}\ln \left( \frac{1}{2}\left(
\tau +c\right) ^{\bar{C}-1}\right) \right]  \label{StM.46} \\
N\left( \tau \right) & =\frac{2}{\sqrt{3kV_{0}}}\exp \left( \frac{\sqrt{%
6k\varepsilon }}{2}\phi \left( \tau \right) \right) .
\end{align}

\subsubsection{The case $K\neq 1$}

\paragraph{\noindent Subcase $K<1$}

In this case the analytic solution of the system of equations (\ref{StM.22}-%
\ref{StM.25}) is \textbf{(}for details see Appendix\textbf{)}%
\begin{align}
a^{3}\left( \tau \right) & =\frac{3}{8}\left( 1-K\right) ^{\left( 1-K\right)
}\left( 1+K\right) ^{\left( 1+K\right) }w\left( \tau \right)
^{2}e^{-2K~p\left( \tau \right) } \\
\phi \left( \tau \right) & =\frac{2}{\sqrt{6k\varepsilon }}\left( \ln \left( 
\frac{\left( 1-K\right) ^{1-K}}{\left( 1+K\right) ^{1+K}}\right) +K\ln \frac{%
8}{3w\left( \tau \right) ^{2}}+2p\left( \tau \right) \right) \\
N\left( \tau \right) & =\frac{2}{\sqrt{3kV_{0}}}\exp \left( \frac{\sqrt{%
6k\varepsilon }}{2}K\phi \left( \tau \right) \right)
\end{align}%
where%
\begin{align}
w\left( \tau \right) & =\sqrt{Ae^{2\omega \tau }-\frac{\bar{E}}{\omega ^{2}}%
+Be^{-2\omega \tau }} \\
p\left( \tau \right) & =\frac{p_{0}\omega }{\sqrt{4\omega ^{4}AB-\bar{E}^{2}}%
}\arctan \left( \frac{2\omega ^{2}Ae^{2\omega \tau }-\bar{E}}{\sqrt{4\omega
^{4}AB-\bar{E}^{2}}}\right)
\end{align}%
and the constants $A,B,p_{0},\bar{C},\bar{E},\omega $ are related with the
constraint 
\begin{equation}
\frac{2\bar{C}}{\omega ^{2}}=-2\omega ^{2}AB-\frac{p_{0}^{2}}{2}+\frac{\bar{E%
}^{2}}{2\omega ^{2}}.
\end{equation}

\paragraph{\noindent Subcase $K>1$}

In this case the analytic solution of the system of equations (\ref{StM.22}-%
\ref{StM.25}) is 
\begin{align}
a^{3}\left( \tau \right) & =\frac{3}{8}\left( K-1\right) ^{\left( 1-K\right)
}\left( 1+K\right) ^{\left( 1+K\right) }\bar{w}\left( \tau \right)
^{2}e^{-2K~\bar{p}\left( \tau \right) } \\
\phi \left( \tau \right) & =\frac{2}{\sqrt{6k\varepsilon }}\left( \ln \left( 
\frac{\left( K-1\right) ^{1-K}}{\left( 1+K\right) ^{1+K}}\right) +K\ln \frac{%
8}{3\bar{w}\left( \tau \right) ^{2}}+2\bar{p}\left( \tau \right) \right) \\
N\left( \tau \right) & =\frac{2}{\sqrt{3kV_{0}}}\exp \left( \frac{\sqrt{%
6k\varepsilon }}{2}K\phi \left( \tau \right) \right)
\end{align}%
where 
\begin{align}
\bar{w}\left( \tau \right) & =\sqrt{\bar{A}e^{2i\bar{\omega}\tau }-\frac{%
\bar{E}}{\bar{\omega}^{2}}+\bar{B}e^{-2i\bar{\omega}\tau }} \\
\bar{p}\left( \tau \right) & =-\frac{i\bar{p}_{0}\bar{\omega}}{\sqrt{4\bar{%
\omega}^{4}\bar{A}\bar{B}-E^{2}}}\arctan \left( \frac{2\bar{\omega}^{2}\bar{A%
}e^{2i\bar{\omega}t}-\bar{E}}{\sqrt{4\bar{\omega}^{4}\bar{A}\bar{B}-\bar{E}%
^{2}}}\right)
\end{align}%
and the constants $\bar{A},\bar{B},\bar{p}_{0},\bar{C},\bar{E},\bar{\omega}$
are related with the constraint%
\begin{equation}
\frac{2\bar{C}}{\bar{\omega}^{2}}=-2\bar{\omega}^{2}\bar{A}\bar{B}+\frac{%
\bar{p}_{0}^{2}}{2}+\frac{\bar{E}^{2}}{2\bar{\omega}^{2}}.
\end{equation}

{\LARGE \ }Our solution reduces to that of \cite{ChimentoTwoSF} if we set $%
\bar{E}=0,$ that is $\mu _{DM}=0$ and to that of \cite{Basilakos} if we set $%
C=0,$\ that is $p_{B}=\mu _{B}=0.$ Finally if we set $C=0,~\bar{E}=0$ we
obtain the solution of \cite{Russo} .

\section{Conclusion}

In order to model all types of matter in the universe we have considered a
mixture of three fluids, a perfect fluid with a linear EoS, a dust for DM
and a scalar field for DE. All fluids are assumed to be minimally
interacting and the background space is assumed to be the flat FRW
spacetime. The available field equations do not suffice to determine the
dynamical system. We need two more assumptions, one which will specify the
scalar field potential and another which will make the system of dynamical
equations solvable. In order to define these conditions we follow two steps.
First we show that the system of the three fluids is equivalent to a two
dimensional dynamical system moving in $M^{2}$ under the constraint $\bar{E}%
= $constant. Then we require that the two dimensional system admits an extra
Lie symmetry. This requirement fixes the potential to be exponential.
Requiring further that the solution is invariant under the Lie symmetry we
obtain a two parameter family of analytic solutions containing a perfect
fluid and a (quintessence of phantom) scalar field with an exponential
potential.

Requiring next that the Lagrangian defining the two dimensional dynamical
system admits an extra Noether symmetry, we obtain a new analytic solution
which involves all three types of matter, that is, stiff matter, scalar
field and dust. It is interesting (see Appendix) that in this case the
dynamical system reduces to the Ermakov Pinney dynamical system which is
known to be integrable \cite{HaasG,LeachKarasu}, hence we are able to obtain
the analytic solution. This solution includes previously found solutions in 
\cite{Basilakos,Russo,ChimentoTwoSF} as special cases.

\textbf{Acknowledgments.} We would like to thank Dr Basilakos S for comments
and suggestions and the referee for useful references. This work has been
partially supported from ELKE (grant 1112) of the University of Athens.

\section*{APPENDIX}

\subsection*{Analytic solutions for $K\neq 1$}

\label{AnalyticAppen}

\subsubsection*{Subcase $K<1$}

We introduce new coordinates $w,p$ with the relations 
\begin{equation}
u=\frac{\ln \left( \frac{\sqrt{3}\left( 1-K\right) }{2\sqrt{2}}we^{p}\right) 
}{1-K},~v=\frac{\ln \left( \frac{\sqrt{3}\left( 1+K\right) }{2\sqrt{2}}%
we^{-p}\right) }{1+K}.  \label{StM.50}
\end{equation}%
Then the Lagrangian (\ref{StM.29}) becomes%
\begin{equation*}
L=\frac{1}{2}w^{\prime 2}-\frac{1}{2}w^{2}p^{\prime 2}+\frac{1}{2}\omega
^{2}w^{2}+\frac{2\bar{C}}{\omega ^{2}}\frac{1}{w^{2}}
\end{equation*}%
where $\omega ^{2}=1-K^{2}.$ This is the Lagrangian for the hyperbolic
Ermakov Pinney dynamical system which is known to be integrable \cite%
{HaasG,LeachKarasu}. The equations of motion are: 
\begin{align}
w^{\prime \prime }+wp^{\prime 2}-\omega ^{2}w+\frac{4C}{\omega ^{2}}\frac{1}{%
w^{3}}& =0  \label{StM.52} \\
p^{\prime \prime }+\frac{2}{w}p^{\prime }w^{\prime }& =0  \label{StM.53}
\end{align}%
The Hamiltonian is:%
\begin{equation}
\bar{E}=\frac{1}{2}w^{\prime 2}-\frac{1}{2}w^{2}p^{\prime 2}-\frac{\omega
^{2}}{2}w^{2}-\frac{2\bar{C}}{\omega ^{2}}\frac{1}{w^{2}}.  \label{StM.54}
\end{equation}%
The solution of the system of equations is%
\begin{align}
w\left( \tau \right) & =\sqrt{Ae^{2\omega \tau }-\frac{\bar{E}}{\omega ^{2}}%
+Be^{-2\omega \tau }}  \label{StM.55} \\
p\left( \tau \right) & =\frac{p_{0}\omega }{\sqrt{4\omega ^{4}AB-\bar{E}^{2}}%
}\arctan \left( \frac{2\omega ^{2}Ae^{2\omega \tau }-\bar{E}}{\sqrt{4\omega
^{4}AB-\bar{E}^{2}}}\right) .  \label{StM.56}
\end{align}%
Using the Hamiltonian we find the constrain%
\begin{equation}
\frac{2\bar{C}}{\omega ^{2}}=-2\omega ^{2}AB-\frac{p_{0}^{2}}{2}+\frac{\bar{E%
}^{2}}{2\omega ^{2}}  \label{StM.57}
\end{equation}%
where $A,B,p_{0}$ are constants.

\subsubsection*{Subcase $K>1$}

In this case we apply the transformation%
\begin{equation}
u=\frac{\ln \left( \frac{\sqrt{3}\left( K-1\right) }{2\sqrt{2}}\bar{w}e^{%
\bar{p}}\right) }{1-K},~v=\frac{\ln \left( \frac{\sqrt{3}\left( 1+K\right) }{%
2\sqrt{2}}\bar{w}e^{-\bar{p}}\right) }{1+K}  \label{StM.65}
\end{equation}%
and the Lagrangian (\ref{StM.29}) becomes%
\begin{equation}
L=-\frac{1}{2}\bar{w}^{\prime 2}+\frac{1}{2}\bar{w}^{2}\bar{p}^{\prime 2}+%
\frac{1}{2}\bar{\omega}^{2}\bar{w}^{2}+\frac{2\bar{C}}{\bar{\omega}^{2}}%
\frac{1}{\bar{w}^{2}}  \label{StM.67}
\end{equation}%
where $\bar{\omega}=K^{2}-1.~$This is the Lagrangian for the harmonic
Ermakov Pinney dynamical system. \ \newline
The solution of the Euler-Lagrange equations is 
\begin{align}
\bar{w}\left( \tau \right) & =\sqrt{\bar{A}e^{2i\bar{\omega}\tau }-\frac{%
\bar{E}}{\bar{\omega}^{2}}+\bar{B}e^{-2i\bar{\omega}\tau }}  \label{StM.71}
\\
\bar{p}\left( \tau \right) & =-\frac{i\bar{p}_{0}\bar{\omega}}{\sqrt{4\bar{%
\omega}^{4}\bar{A}\bar{B}-E^{2}}}\arctan \left( \frac{2\bar{\omega}^{2}\bar{A%
}e^{2i\bar{\omega}t}-\bar{E}}{\sqrt{4\bar{\omega}^{4}\bar{A}\bar{B}-\bar{E}%
^{2}}}\right)  \label{StM.72}
\end{align}%
where $\bar{A},\bar{B},\bar{p}_{0}$ are constants which are related with the
constraint equation%
\begin{equation}
\frac{2\bar{C}}{\bar{\omega}^{2}}=-2\bar{\omega}^{2}\bar{A}\bar{B}+\frac{%
\bar{p}_{0}^{2}}{2}+\frac{\bar{E}^{2}}{2\bar{\omega}^{2}}.  \label{StM.73}
\end{equation}


\begin{thebibliography}{99}
\bibitem{Weinberg89} Weinberg S 1989 Rev. Mod. Phys. \textbf{61} 1

\bibitem{Peebles03} Peebles P J and Ratra B 2003 Rev. Mod. Phys. \textbf{75}
559

\bibitem{Pad03} Padmanabhan T 2003 Phys. Rept. \textbf{380} 235

\bibitem{Peri08} Perivolaropoulos L. 2008 [arxXiv.0811.4684]

\bibitem{Ratra88} Ratra B and Peebles P J E 1988 Phys. Rev D. \textbf{37}
3406

\bibitem{Lambda4} Overduin J M and Cooperstock F I 1998 Phys. Rev. D \textbf{%
58} 043506

\bibitem{Bas09c} Basilakos S Plionis M and Sol\`{a} S 2009 Phys. Rev. D 
\textbf{80} 3511

\bibitem{Linder2004} Linder E V 2004 Phys.\ Rev.\ Lett.\ \textbf{70} 023511

\bibitem{LSS08} Lima J A S Silva F E and Santos R C 2008 Class. Quantum
Grav. \textbf{25} 205006

\bibitem{Brookfield2005td} Brookfield A W et.al 2006 Phys. Rev. Lett. 
\textbf{96} 061301

\bibitem{Basilakos} Basilakos S Tsamparlis M and Paliathanasis A 2011 Phys.
Rev. D \textbf{83}, 103512

\bibitem{Cap09} Capozziello S Piedipalumbo E Rubano C and Scudellaro P 2009
Phys. Rev. D \textbf{80} 104030

\bibitem{deRitis} de Ritis R Marmo G et al 1990 Phys Rev D \textbf{42} 1091

\bibitem{RubanoSFQ} Rubano C and Scudellaro P 2002 Gen. Relativ. Gravit. 
\textbf{34} 307

\bibitem{YiZhang} Zhang Y Gong Y G and Zhu Z\ H 2010 Phys. Lett. B \textbf{%
688} 13

\bibitem{Tsam10} Tsamparlis M and Paliathanasis A 2011~J. Phys. A: Math. and
Theor. \textbf{44} 175202

\bibitem{Olver} Olver P J 1986\ Applications of Lie Groups to Differential
Equations Springer - Verlag N.Y.

\bibitem{StephaniB} Stephani H 1989 Differential Equations: Their Solutions
using Symmetry Cambridge University Press

\bibitem{Aminova 1995} Aminova A V 1995 Sbornik Mathematics \textbf{186} 1711

\bibitem{HaasG} Haas F and Goedert J 2001 Phys. Lett. A \textbf{279} 181

\bibitem{LeachKarasu} Leach P G L and Karasu A 2002 \ J. Nonlinear Math.
Phys. \textbf{9} 475\qquad

\bibitem{Russo} Russo J G 2004 Phys. Lett. B \textbf{600} 185

\bibitem{Bertacca} Bertacca D Matarrese S and Pietroni M 2007 Mod. Phys.
Lett. A \textbf{22} 2893

\bibitem{MuslimovSF} Muslimov A G 1990 Class. Quantum Grav. \textbf{7} 231

\bibitem{MendezExactSF} Mendez V 1996 Class. Quantum Grav. \textbf{13} 3229

\bibitem{EllisExactSF} Ellis G F R and Madsen M S 1991 Class. Quantum Grav. 
\textbf{8} 667

\bibitem{ChimentoTwoSF} Chimento L P 1998 Class. Quantum Grav. \textbf{15}
965

\bibitem{HalliwellSF} Halliwell J J 1987 Phys. Lett. B. \textbf{185} 341

\bibitem{Easther} Easther R 1993 Class. Quantum Grav. \textbf{10} 2203

\bibitem{Barrow} Barrow J D 1993 Class. Quantum Grav. \textbf{10} 279

\bibitem{ChimentoLPC} Chimento L P and Cossarini A E 1994 Class. Quantum
Grav. \textbf{11} 1177

\bibitem{ChimentoJacubi} Chimento L P and Jakubi A S 1996 Int. J. of Mod.
Phys. D \textbf{5} 71

\bibitem{CapNojiri} Capozziello S Nojiri S and Odintsov S D 2006 Phys. Lett.
B \textbf{632} 597

\bibitem{NojiriOdi} Nojiri S and Odintsov S D 2006 Gen. Relativ. Gravit. 
\textbf{38} 1285

\bibitem{Cap97} Capozziello S de Ritis R and Marino A 1997 Class. Quantum
Grav. \textbf{14} 3259

\bibitem{Sanyal} Sanyal A K Modak B Rubano C and \ Piedipalumbo E 2005 Gen.
Relativ. Grav. \textbf{37} 407

\bibitem{Cap07} Capozziello S Stabile A and Troisi A 2007 Class. Quantum
Grav. \textbf{\ 24} 2153

\bibitem{RoshanPalatiniNoether} Roshan M and Shojai F 2008 Phys. Lett. B 
\textbf{668} 238

\bibitem{BonannoGLNoether} Bonanno A Esposito\ G Rubano C and Scudellaro P
2007 Gen. Relativ. Gravit. \textbf{39} 189

\bibitem{VakiliFr} Vakili B 2008 Phys. Lett. B \textbf{664} 16

\bibitem{Motavali} Motavali H Capozziello S and Rowshan Almeh Jog M 2008
Phys. Lett. B \textbf{666} 10

\bibitem{MahomedTF} Jamil M Mahomed F M and Moment D 2011 Phys. Lett. B 
\textbf{702} 315
\end{thebibliography}
\end{document}